\documentstyle[aps,prl,epsfig]{revtex}

\newcommand{\Dslash}{{D \hskip -6pt /}}

\bibstyle{unsrt}
\begin{document}
\draft

\title{Borel Summation of the Derivative Expansion and Effective Actions}
\author{Gerald V. Dunne and Theodore M. Hall}
\address{Department of Physics, University of Connecticut, Storrs CT
06269, USA}

\maketitle
\vskip .5cm

\begin{abstract}
We give an explicit demonstration that the derivative expansion of the QED effective action is a divergent but Borel summable asymptotic series, for a particular inhomogeneous background magnetic field. A duality transformation $B\to iE$ gives a non-Borel-summable perturbative series for a time dependent background electric field, and Borel dispersion relations yield the non-perturbative imaginary part of the effective action, which determines the pair production probability. Resummations of leading Borel approximations exponentiate to give perturbative corrections to the exponents in the non-perturbative pair production rates. Comparison with a WKB analysis suggests that these divergence properties are general features of derivative expansions and effective actions.
\end{abstract}

\section{Introduction}

The effective action plays a central role in quantum field theory. Here we
consider the one-loop effective action in quantum electrodynamics (QED) for electrons in the presence of a background electromagnetic field:
\begin{equation}
S=-i \ln \det (i\Dslash-m)=-\frac{i}{2}\ln \det(\Dslash^2+m^2)
\label{action}
\end{equation}
where $\Dslash =\gamma^\nu\left(\partial_\nu+ie A_\nu\right)$, and $A_\nu$ is a
fixed classical gauge potential with field strength tensor
$F_{\mu\nu}=\partial_\mu A_\nu- \partial_\nu A_\mu$. When the background is a static magnetic field, the effective action $S$ is equal to minus the effective energy of the electrons in that background; and when the background is an electric field, $S$ has an imaginary part which determines the pair-production rate for electron-positron pair creation \cite{euler,schwinger,stone1,greiner}. For a uniform background field strength $F_{\mu\nu}={\rm constant}$, the effective action $S$ can be computed exactly \cite{euler,schwinger,weisskopf,nikishov,matt}. For more general backgrounds, with $F_{\mu\nu}$ {\it not} constant, the situation is more complicated. One standard approach is to make a ``derivative expansion'' \cite{aitchison,lee,daniel,shovkovy,schubert} (or ``gradient expansion'' \cite{stone2}) which is a formal perturbative expansion in increasing numbers of derivatives of $F_{\mu\nu}$ :
\begin{equation}
S=S^{(0)}[F]+S^{(2)}[F,(\partial F)^2]+\dots
\label{derivative}
\end{equation}
In this paper we address two questions concerning the QED effective action $S$ in a nonuniform background. First, we consider the convergence/divergence properties of the perturbative derivative expansion in (\ref{derivative}). Second, we ask how such a perturbative expansion can lead to corrections to Schwinger's nonperturbative pair-production rate (computed for a constant background) when there are inhomogeneities in the background electric field. We can answer these questions precisely by considering some exactly solvable cases with special inhomogeneous backgrounds \cite{dh1,dh2,ted}. The derivative expansion is found to be a divergent asymptotic series, and the rate of divergence at high orders can be used to compute the corresponding non-perturbative imaginary part of the effective action when the background is a time-dependent electric field. This divergence of the derivative expansion is not a bad thing; it is completely analogous to generic behavior that is well known in perturbation theory in both quantum field theory and quantum mechanics. For example, Dyson \cite{dyson} argued that QED perturbation theory cannot be convergent as an expansion in the fine structure constant $\alpha$, because the theory is unstable when $\alpha$ is negative. (The divergent and asymptotic nature of field theoretic perturbation theory was also found in scalar $\phi^3$ theories \cite{hurst}). In fact, the Euler-Heisenberg-Schwinger QED effective action in a constant background, and the QED effective action in the inhomogeneous backgrounds discussed in this paper, provide explicit demonstrations of Dyson's argument. 

Our analysis of the convergence/divergence properties of the QED effective action uses Borel summation \cite{hardy,carl}, a mathematical tool that can be used to relate the rate of divergence of high orders of perturbation theory to non-perturbative decay and tunneling rates, thereby providing a bridge between perturbative and non-perturbative physics. Other well known explicit cases of this connection appear in quantum mechanical  
examples such as the anharmonic oscillator \cite{bw} and the Stark effect \cite{simon}, and in quantum field theory in semi-classical analyses of scalar field theories \cite{lipatov} and asymptotic estimates of large orders of QED perturbation theory \cite{ipz}. For an excellent review of a broad range of examples, see Ref. \cite{zinn}. Typically one finds that in a stable situation ({\it i.e.}, no tunneling or decay processes) perturbation theory is divergent, with expansion coefficients that alternate in sign and grow factorially in magnitude. On the other hand, in an unstable situation, perturbation theory is generally divergent with coefficients that grow factorially in magnitude but do not alternate in sign. Borel summation is an approach to the summation of divergent series that makes physical sense out of these two different types of behavior. We shall see that the divergence of the derivative expansion can be understood naturally in this Borel framework.

In addition to these theoretical considerations of understanding the connections between the perturbative derivative expansion and non-perturbative pair-production rates, another motivation for this work is provided by 
the attempt to observe electron-positron pair creation due to QED vacuum effects in the presence of strong electric fields. Schwinger's constant field pair-production rate is far too small to be accessible with present electric field strengths. However, the constant field approximation is somewhat unrealistic, and so one can ask how this rate is modified by a time variation of the electric background. For sinusoidal time variation Br\'ezin and Itzykson found a WKB result with fairly weak frequency dependence \cite{brezin}, while Balantekin {\it et al} have applied group theory and uniform WKB to electric backgrounds with more general time dependence \cite{baha}. The QED effective action has recently been computed \cite{dh2} as a closed form (single integral) expression for the particular time-dependent electric background with $E(t)=E\, {\rm sech}^2(t/\tau)$, and in this paper we consider the numerical implications of this result for pair-production rates in such a background. It is important to note that another related approach to observing pair-production is to use highly relativistic electrons as intermediate states, as has been done in recent experiments \cite{burke,melissinos}.

Finally, we note that the derivative expansion (\ref{derivative}) is an example of an {\it effective field theory} expansion \cite{manohar}, such as is used in operator product expansions \cite{shifman} and chiral perturbation theory \cite{donoghue}. In the effective field theory approach, the mass $m$ of the electrons (which are `integrated out' in the one-loop approximation) sets an energy scale, and the physics at energies ${\cal E}\ll m$ should be described by a low energy effective action with the formal expansion
\begin{equation}
S=m^4\, \sum_n a_n\, {O^{(n)}\over m^n}
\label{eff}
\end{equation}
where $O^{(n)}$ is an operator of dimension $n$. For the case of the QED effective action, the simplest way to produce higher-derivative operators in such an expansion is to take higher powers of the field strength $F$. Thus, even the leading ({\it i.e.} constant background) term $S^{(0)}[F]$ in the derivative expansion (\ref{derivative}) is itself an effective field theory expansion of the form (\ref{eff}). This is simply the Euler-Heisenberg effective action which is a perturbative series expansion in powers of $e^2 F^2/m^4$. This series is known to be divergent \cite{og,olesen,popov}, and there are important physical consequences of this divergence, as we review below. Another way to produce higher dimension operators in the expansion (\ref{eff}) is to include derivatives of $F$, with each derivative balanced by an inverse power of $m$. This is what is done in the derivative expansion (\ref{derivative}). Thus, we can view the derivative expansion (\ref{derivative}) as a `double' series expansion, both in powers of $F$ and in derivatives of $F$. In this paper we study the divergence properties of such an expansion. It has been suggested, based on the behavior of the constant field case \cite{zhitnitsky}, that the effective field theory expansion (\ref{eff}) is generically divergent. Here we provide an explicit demonstration of this divergence for inhomogeneous background fields. 

For energies well below the scale set by the fermion mass $m$, the divergent nature of the effective action is not important, as the first few terms provide an accurate approximation. However, the divergence properties do become important when the external energy scale approaches the fermion mass scale $m$, and/or when the inhomogeneity scale becomes short compared to a characteristic scale of the system. The divergence is also important for understanding how the non-perturbative imaginary contributions to the effective action arise from {\it real} perturbation theory.

In Section II we review briefly the mathematical technique of Borel summation, and in Section III we apply this to the Euler-Heisenberg-Schwinger
constant-background effective action. Section IV gives the Borel summation analysis of the particular exactly solvable cases with inhomogeneous magnetic and electric backgrounds. In Section V we show how this is related to a WKB analysis, and Section VI contains some concluding remarks.

\section{Brief Review of Borel Summation}

In this section we review briefly the basics of Borel summation \cite{hardy,carl}. Consider an asymptotic series expansion of some function $f(g)$
\begin{eqnarray}
f(g)\sim \sum_{n=0}^\infty \, a_n\, g^n\qquad\qquad (g\to0^+)
\label{exp}
\end{eqnarray}
where $g>0$ is a (small) dimensionless perturbation expansion parameter and the $a_n$ are real coefficients. 
In an extremely broad range of physics applications \cite{zinn,popov} it has 
been found that perturbation theory leads not to a convergent series but to a
divergent series like (\ref{exp}) in which the expansion coefficients $a_n$ have
large-order behavior of the form
\begin{eqnarray}
a_n\sim (-1)^n \alpha^n \Gamma(\beta n+\gamma) \left[1+O\left(\frac{1}{n}\right)\right] \qquad\qquad (n\to\infty)
\label{general}
\end{eqnarray}
for some real constants $\alpha$, $\beta>0$, and $\gamma$. When $\alpha>0$, the perturbative expansion coefficients $a_n$ alternate in sign and their magnitude grows factorially. Borel summation is a particularly useful approach for this case of a divergent, but alternating series. We shall see below that non-alternating series must be treated somewhat differently.

Consider, for example, the asymptotic series (\ref{exp}) with 
$a_n=(-1)^n \alpha^n n!$, and $\alpha>0$. This series is clearly divergent for any value of the expansion parameter $g$. Borel summation of this divergent series can be motivated by the following formal procedure. Write
\begin{eqnarray}
n!= \int_0^\infty \, ds \, s^n\, e^{-s}
\label{fact}
\end{eqnarray}
and then formally interchange the order of summation and integration, to yield
\begin{eqnarray}
f(g)\sim \frac{1}{\alpha g}\, \int_0^\infty \,ds\, \left({1\over 1+s}\right) \, 
\exp\left[- \frac{s}{\alpha g}\right] \qquad\qquad (g\to 0^+)
\label{borel}
\end{eqnarray}
This integral is convergent for all $g >0$, and so can be used to {\it define} the sum of the divergent series $\sum_{n=0}^\infty (-1)^n \alpha^n n!g^n$. To be more precise, the formula (\ref{borel}) should be read from right to left: for $g\to 0^+$, we can use Laplace's method \cite{carl} to make an asymptotic expansion of the integral, and we obtain the asymptotic series in (\ref{exp}) with expansion coefficients $a_n=(-1)^n \alpha^n n!$. 

The Borel integral (\ref{borel}) can be analytically continued off the $g>0$ axis and in this case is in fact \cite{hardy,carl} an analytic function of $g$ in the cut $g$ plane: $|{\rm arg}(g)|<\pi$. Thus, we can use a simple dispersion relation (using the discontinuity across the cut along the negative $g$ axis) to {\it define} the imaginary part of $f(g)$ for negative values of the expansion parameter:
\begin{eqnarray}
{\rm Im} f(-g)\sim\frac{\pi}{\alpha g}\exp[-\frac{1}{\alpha g}] \qquad\qquad (g\to 0^+)
\label{imag}
\end{eqnarray}
Note, of course, that an alternating series with negative $g$ is the same as a 
{\it non-alternating} series with positive $g$. If the expansion coefficients in (\ref{exp}) are {\it non-alternating} (with $g>0$) then the situation is very different, both physically and mathematically. Formal application of Borel summation yields
\begin{eqnarray}
\sum_{n=0}^\infty \alpha^n n! \, g^n \sim \frac{1}{\alpha g}\int_0^\infty ds\,\left(\frac{1}{1-s}\right)\exp\left[-\frac{s}{\alpha g}\right]  \qquad\qquad (g\to 0^+)
\label{non}
\end{eqnarray}
This Borel integral has a pole on the integration contour. However, a
principal parts prescription for such a pole gives an imaginary part in
agreement with (\ref{imag}). This imaginary contribution is non-perturbative (it clearly does not have an expansion in positive powers of $g$) and has important physical consequences. Generically, the Borel technique signals the possible presence of such non-perturbative physics if the perturbative expansion coefficients grow rapidly (factorially) in magnitude and are
non-alternating. The associated dispersion relations provide a bridge between the perturbative physics [i.e. the $a_n$'s] and the non-perturbative imaginary parts [i.e. the $\exp(-\frac{1}{\alpha g})$ factors]. We will see explicit examples of this below.

We should note at this point that for a general divergent series these Borel dispersion relations may be complicated by the appearance of additional poles and/or cuts in the complex $g$ plane. Physically interesting poles, known as renormalons, are indeed found in certain resummations of perturbation theory for both QED and QCD \cite{thooft,zakharov,beneke}. Here, for the one-loop QED effective action in a fixed external background we find that we do not encounter such poles.

The Borel summation construction discussed above for the case $a_n=(-1)^n \alpha^n n!$, generalizes in the obvious way to the case 
where the perturbative coefficients are 
$a_n= (-1)^n \alpha^n \Gamma(\beta n+\gamma)$, which corresponds to the leading-order growth indicated in (\ref{general}):
\begin{eqnarray}
f(g)&\sim& \sum_{n=0}^\infty (-1)^n \alpha^n \Gamma(\beta n+\gamma) g^n \nonumber\\
&\sim& \frac{1}{\beta}\, \int_0^\infty \frac{ds}{s} \, \left(\frac{1}{1+s}\right)
\left(\frac{s}{\alpha g}\right)^{\gamma/\beta}\, 
\exp\left[-\left(\frac{s}{\alpha g}\right)^{1/\beta}\right] \qquad\qquad (g\to 0^+)
\label{genborel}
\end{eqnarray}
The corresponding imaginary part for negative values of the expansion parameter is
\begin{eqnarray}
{\rm Im} f(-g)\sim\frac{\pi}{\beta}\left(\frac{1}{\alpha g} \right)^{\gamma/\beta} 
\exp\left[-\left(\frac{1}{\alpha g}\right)^{1/\beta}\right]
 \qquad\qquad (g\to 0^+)
\label{genimag}
\end{eqnarray}
Notice that the parameter $\beta$ affects the exponent, while the combination $\frac{\gamma}{\beta}$ is important for the prefactor. 

These last two formulas, the Borel integral (\ref{genborel}) and the Borel dispersion relation (\ref{genimag}), will be used repeatedly below.

\section{Euler-Heisenberg-Schwinger Effective Action}

Now consider applying this Borel summation machinery to QED effective 
actions. Effective actions can be expanded perturbatively in terms of the coupling constant $e$, and also in terms of derivatives of the background field strength $F_{\mu\nu}$. To begin, we review the well-known Euler-Heisenberg-Schwinger effective action which corresponds to a uniform background field strength; thus the only expansion is in terms of the perturbative coupling constant $e$. We consider first a magnetic background, and then we consider an electric background. For a uniform background {\it magnetic} field of strength $B$, the exact renormalized effective action can be expressed as a `proper-time' integral \cite{schwinger}
\begin{equation}
S=-\frac{e^2 B^2 L^3 T}{8\pi ^{2}} \int_{0}^{\infty} \frac{ds}{s^{2}}\; 
(\coth s-\frac{1}{s}-\frac{s}{3})\,e^{-m^2s/(eB)}
\label{proper}
\end{equation}
The $\frac{1}{s}$ term is a subtraction of the zero field ($B=0$) effective action, while the $\frac{s}{3}$ subtraction corresponds to a logarithmically divergent charge renormalization \cite{schwinger}. The $L^3 T$ factor is the space-time volume factor. It is straightforward to develop, for small $\frac{eB}{m^2}$, an asymptotic expansion of this integral:
\begin{eqnarray}
S&\sim& -\frac{2 e^2 B^2 L^3 T}{\pi^2}\frac{e^2 B^2}{m^4} \sum_{n=0}^\infty 
\frac{{\cal B}_{2n+4}} {(2n+4)(2n+3)(2n+2)}\left(\frac{2eB}{m^2}\right)^{2n}
\nonumber\\
&=&{m^4 L^3 T\over \pi^2}\left[ {1\over 360}\left(\frac{eB}{m^2}\right)^4 -
{1\over 630} \left(\frac{eB}{m^2}\right)^6 +
{1\over 315} \left(\frac{eB}{m^2}\right)^8 - \dots \right] \qquad\qquad (\frac{eB}{m^2}\to 0^+)
\label{eh}
\end{eqnarray}
Here the ${\cal B}_{2n}$ are Bernoulli numbers \cite{bern}. The perturbative series (\ref{eh}) is the Euler-Heisenberg \cite{euler,weisskopf} perturbative expression for the QED effective action in a uniform magnetic background $B$. It is an expansion in powers of the coupling $e$, with the $n^{\rm th}$ power of $e$ being associated with a one-fermion-loop diagram with $n$ external photon lines [we have not included the divergent ${\rm O}(e^2)$ self-energy term as it contributes to the bare action by charge renormalization]. Note that only even powers of $eB$ appear in the perturbative expansion (\ref{eh}). This is due to charge conjugation invariance (Furry's theorem). The expansion (\ref{eh}) is also an expansion in inverse powers of $m^2$, as is familiar for an effective field theory action (\ref{eff}), with the higher dimensional operators in the expansion simply being higher powers of $B^2$. 

The Euler-Heisenberg perturbative effective action (\ref{eh}) is not a convergent series. Rather, it is an asymptotic series of the form (\ref{exp}) with expansion parameter
\begin{eqnarray}
g=\frac{4 e^2 B^2}{m^4}
\label{param}
\end{eqnarray}
The expansion coefficients in (\ref{eh}) alternate in sign [because: ${\rm sign}({\cal B}_{2n})=(-1)^{n+1}$], and grow factorially in magnitude:
\begin{eqnarray}
a_n&=&\frac{{\cal B}_{2n+4}} {(2n+4)(2n+3)(2n+2)}\nonumber\\
&\sim& (-1)^{n+1}\frac{2}{(2\pi)^{2n+4}}\Gamma(2n+2) \qquad\qquad ,\,n\to\infty
\label{growth}
\end{eqnarray}
The growth of these coefficients is of the form indicated in the example in  (\ref{general}), with $\alpha=\frac{1}{4\pi^2}$, and $\beta=\gamma=2$. Therefore, the Euler-Heisenberg series (\ref{eh}) is an example of a divergent but Borel summable series. If we keep just the leading large-n behavior for the coefficients $a_n$ indicated in (\ref{growth}), then we can immediately read off from the Borel summation formula (\ref{genborel}) the leading-order Borel approximation for the sum of the divergent series (\ref{eh}):
\begin{eqnarray}
S_{\rm leading}\sim {e^2 B^2 L^3 T\over 4\pi^6} \int_0^\infty \, ds 
\left({s\over 1+s^2/\pi^2}\right) e^{-m^2 s/(eB)} \qquad\qquad (\frac{eB}{m^2}\to 0^+)
\label{firstborel}
\end{eqnarray}
It is straightforward to evaluate this integral numerically, and one finds approximately $10 - 15\%$ agreement with the exact answer (\ref{proper}) even when the perturbative expansion parameter $g$ is as large as 50, as is shown in Figure 1. (Note that (\ref{growth}) suggests it is perhaps more `natural' to take the expansion parameter to be $g/(2\pi)^2 \approx g/40$, so we have plotted the leading Borel approximation (\ref{firstborel}) for $g$ up to $50$.)

\begin{figure}[htb]
\centering{\epsfig{file=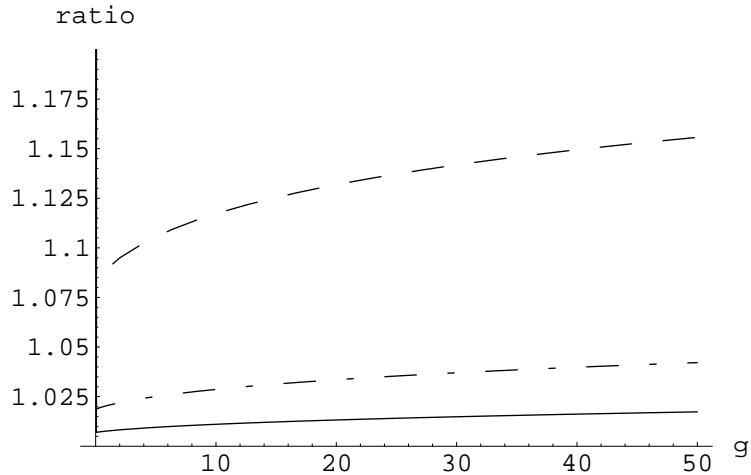, width = 4in, height=2.5in}}
\vskip .5cm
\caption{This figure plots, as a function of the dimensionless expansion parameter $g$ defined in (\protect{\ref{param}}), the ratio of the exact effective action (\protect{\ref{proper}}) to the Borel summation approximation in (\protect{\ref{fullborel}}). The dashed line refers to just the leading Borel approximation in (\protect{\ref{firstborel}}), while the dot-dash line refers to taking the first two terms in the expansion (\protect{\ref{fullborel}}), and the solid line refers to taking the first three terms in (\protect{\ref{fullborel}}).}
\end{figure}

But in this case we can do much better, because we are in the unusual situation of knowing the {\it exact} perturbative expansion coefficients $a_n$ {\it for all} $n$ (not simply their leading-order growth): 
\begin{eqnarray}
a_n=\frac{{\cal B}_{2n+4}} {(2n+4)(2n+3)(2n+2)}&=&
(-1)^{n+1}\frac{2}{(2\pi)^{2n+4}}\Gamma(2n+2)\zeta(2n+4)\nonumber\\
&=& 2 (-1)^{n+1}\Gamma(2n+2) \sum_{k=1}^\infty 
\frac{1}{(2\pi k)^{2n+4}}
\label{zeta}
\end{eqnarray}
For each $k$ in this sum, the coefficient is once again of the leading form in (\ref{general}), with $\alpha=\frac{1}{4\pi^2 k^2}$, and $\beta=\gamma=2$. Thus, we can apply the Borel summation formula (\ref{genborel}) directly to yield
\begin{eqnarray}
S\sim {e^2 B^2 L^3 T\over 4\pi^6} \sum_{k=1}^\infty \, \int_0^\infty ds\,  \left({s\over k^2(k^2+s^2/\pi^2)}\right)\, e^{-m^2 s/(eB)} \qquad\qquad (\frac{eB}{m^2}\to 0^+)
\label{fullborel}
\end{eqnarray}
The sum over $k$ gives successive corrections to the leading Borel approximation in (\ref{firstborel}). The contributions with one, two and three terms are plotted in Figure 1, each compared to the exact result (\ref{proper}). Note that only three terms are needed to obtain $1\%$ accuracy, even when the expansion parameter $g$ is as large as 50. In fact, the expansion \cite{gradshteyn}
\begin{eqnarray}
{\rm coth}(s)-\frac{1}{s}=\frac{2s}{\pi^2}\sum_{k=1}^\infty 
{1\over k^2+ s^2/\pi^2}
\label{coth}
\end{eqnarray}
together with the fact that $\zeta(2)\equiv \sum_{k=1}^\infty 1/k^2=\pi^2/6$, shows that the Borel integral (\ref{fullborel}) agrees precisely with the Schwinger proper-time result (\ref{proper}). That is, Schwinger's formula (\ref{proper}) can be viewed as the Borel sum of the (divergent) Euler-Heisenberg perturbative series (\ref{eh}). Or, in other words, the Euler-Heisnberg perturbative series (\ref{eh}) can be obtained by an asymptotic expansion of Schwinger's integral representation formula (\ref{proper}), when $\frac{eB}{m^2}$ is small.

To get a sense of the size of the expansion parameter (\ref{param}) appearing in the Euler-Heisenberg series, it is instructive to re-instate factors of $\hbar$ and $c$:
\begin{eqnarray}
g=4\left({\hbar\frac{e B}{mc}\over m  c^2}\right)^2 = 4\left(\frac{\hbar/(mc)}{\sqrt{\hbar c/(eB)}}\right)^4
\label{realparam}
\end{eqnarray}
The first equality in (\ref{realparam}) expresses $g$ in terms of the square of the ratio of the cyclotron energy $\hbar \omega_c$ to the electron rest mass energy $mc^2$, while the second equality expresses it in terms of the fourth power of the ratio of the electron Compton wavelength $\frac{h}{mc}$ to the ``magnetic length'' scale $\sqrt{\frac{\hbar c}{eB}}$ set by the magnetic field. The critical magnetic field strength at which the dimensionless parameter $g$ in (\ref{realparam}) is order 1 is $B_c=\frac{1}{2}\frac{m^2c^3}{e\hbar}\sim 10^{13}$ gauss. This is well above currently available laboratory static magnetic field strengths, which are approximately $10^5-10^6$ gauss, in which case $g\sim 10^{-16}-10^{-14}$ is {\it extremely} small. However, the critical field $B_c$ is comparable to the scale of magnetic field strengths observed in astrophysical objects such as supernovae and neutron stars which can have magnetic fields of the order of $10^{15}$ gauss \cite{astro}.

Now consider the Euler-Heisenberg-Schwinger effective action in a uniform background {\it electric} field of strength $E$, instead of the uniform magnetic background $B$. Perturbatively, the only difference is that $B^2$ is replaced by $-E^2$, which amounts to changing the sign of the expansion parameter $g$ in (\ref{param}). Therefore, in a uniform electric background, the Euler-Heisenberg perturbative effective action (\ref{eh}) becomes a 
{\it non-alternating} series 
\begin{eqnarray}
S\sim -\frac{2 e^2 E^2 L^3 T}{\pi^2}\frac{e^2 E^2}{m^4} \sum_{n=0}^\infty 
\frac{(-1)^n {\cal B}_{2n+4}} {(2n+4)(2n+3)(2n+2)}\left(\frac{2eE}{m^2}\right)^{2n} \qquad\qquad (\frac{eE}{m^2}\to 0^+)
\label{ehe}
\end{eqnarray}
(Recall that ${\rm sign}({\cal B}_{2n+4})=(-1)^{n+1}$, so $(-1)^n {\cal B}_{2n+4}$ is non-alternating.) This series is clearly divergent and since the coefficients are non-alternating, it is not Borel summable. Nevertheless, using the Borel dispersion relations we can extract the imaginary part of the effective action. If we keep just the leading large-$n$ growth (\ref{growth}) of the expansion coefficients, then we can immediately read off from the Borel dispersion relation result (\ref{genimag}) the leading behavior of the imaginary part of the effective action in the electric background:
\begin{eqnarray}
{\rm Im} S_{\rm leading} \sim L^3 T\, \frac{e^2 E^2}{8\pi^3} \exp\left[-\frac{m^2\pi}{eE}\right] \qquad\qquad (\frac{eE}{m^2}\to 0^+)
\label{leadingimag}
\end{eqnarray}
This imaginary part has direct physical significance - it gives half the
electron-positron pair production rate in the uniform electric field $E$ \cite{schwinger}. Actually, as in the magnetic case, we can do better than just the leading behavior (\ref{leadingimag}). Combining the expansion coefficients (\ref{zeta}) with the Borel dispersion formula (\ref{genimag}) we immediately find
\begin{eqnarray}
{\rm Im} S\sim L^3 T\, \frac{e^2 E^2}{8\pi^3}\sum_{k=1}^\infty \frac{1}{k^2} \,
\exp\left[-\frac{m^2\pi k}{eE}\right] \qquad\qquad (\frac{eE}{m^2}\to 0^+)
\label{fullimag}
\end{eqnarray}
which is Schwinger's classic result \cite{schwinger}. 

Note that in the electric case the relevant small dimensionless parameter is [compare with (\ref{realparam})]
\begin{eqnarray}
{eE\hbar/(mc)\over  mc^2}
\label{elecparam}
\end{eqnarray}
which is (up to a factor of $\pi$) the ratio of the work done by the electric field $E$ accelerating a particle of charge $e$ through an electron Compton wavelength, to the energy required for pair production. For typical electric fields this is a very small number, so the exponential factors in (\ref{leadingimag}) and (\ref{fullimag}) are extremely small. The critical electric field at which the non-perturbative factors become significant is $E_c=\frac{m^2c^3}{e\hbar}\sim 10^{16}~ Vcm^{-1}$. This is still several orders of magnitude beyond the field obtainable in current lasers \cite{melissinos}.

To conclude this section, we stress that this constant-field case provides an explicit example of Dyson's argument \cite{dyson} that QED perturbation theory cannot be convergent as a series in the fine structure constant $\alpha=\frac{e^2}{4\pi}$, because this would mean that the stable vacuum, with $\alpha$ positive, is smoothly connected to the unstable vacuum, with $\alpha$ negative, at least in a small neighborhood of the origin. The perturbative Euler-Heisenberg series in (\ref{eh}) and (\ref{ehe}) are expansions in powers of $e^2B^2/m^4$ and $e^2E^2/m^4$, respectively. Changing from a magnetic background to an electric background involves replacing $e^2B^2$ with $-e^2E^2$, which amounts to changing the sign of $e^2$ (i.e., the fine structure constant), since $e$ always appears as $eB$ or $eE$. If the Euler-Heisenberg perturbative series were convergent, then the change from $e^2B^2$ to $-e^2E^2$ would not produce any non-perturbative imaginary part in the effective action. Thus there would be no pair production and we would miss the genuine physical instability of the QED vacuum in an external electric field. Hence the Euler-Heisenberg perturbative series {\it cannot} be convergent.

\section{Solvable Inhomogeneous Backgrounds}

So far, we have re-phrased well-known QED results in the language of Borel summation. Now we turn to the main point of this paper, which is to go beyond the Euler-Heisenberg-Schwinger constant field results for the QED effective action. Perturbatively, this leads to a derivative expansion (\ref{derivative}) which is a formal expansion in increasing numbers of derivatives of the background field strength
\begin{eqnarray}
S=S^{(0)}[F_{\mu\nu}] + S^{(2)}[F_{\mu\nu},\partial_\mu F_{\nu\rho}]+ \dots
\label{der}
\end{eqnarray}
where $S^{(0)}$ involves no derivatives of the background field strength $F_{\mu\nu}$, while the first correction $S^{(2)}$ involves two derivatives of the field strength, and so on. The increasing powers of derivatives are balanced by increasing powers of $1/m$.

Unfortunately, it is very difficult to say anything precise about the convergence or divergence of such a derivative expansion because it is not an actual series, as in (\ref{exp}), in terms of a dimensionless expansion parameter. For a general background there is a rapid proliferation of the number of independent terms with a given number of derivatives of the field strength (see \cite{lee,shovkovy} for the first order and \cite{schubert} for higher orders). Even a first order derivative expansion calculation is quite non-trivial. Moreover, for a general background field, it is extremely difficult to estimate and compare the magnitude of the various terms in the derivative expansion (\ref{der}). So a perturbative analysis to high orders in a derivative expansion appears prohibitively difficult for a general background field strength. This makes it difficult to reconcile a perturbative derivative expansion calculation with the calculation of the non-perturbative imaginary part of the effective action for an electric background. We explore this question below.

As a first step towards overcoming these obstacles, we can consider restricted classes of special backgrounds for which the derivative expansion reduces to a manageable form. The QED effective action has recently been computed {\it exactly} for either (not both) of the following special inhomogeneous background magnetic \cite{dh1} and electric fields \cite{dh2}:
\begin{eqnarray}
\vec{B}(x)&=&\vec{B}\, {\rm sech}^2(\frac{x}{\lambda})\nonumber\\
\vec{E}(t)&=&\vec{E}\, {\rm sech}^2(\frac{t}{\tau})
\label{inhom}
\end{eqnarray}
It has of course long been known that the Dirac equation is exactly solvable for such backgrounds, a fact that has permitted many authors to study the QED effective action in these backgrounds \cite{naro,cornwall,chodos}. The new feature of \cite{dh1,dh2} is that all the momentum traces have been performed so that the effective action is expressed as a simple integral representation [involving a {\it single} integral], as in Schwinger's classic result (\ref{proper}) for the uniform background field. This then permits the expansion of the effective action as a true series, whose convergence/divergence properties can be studied in detail. 

On the right side of (\ref{inhom}), $\vec{B}$ and $\vec{E}$ are constant vectors. Thus, $\vec{B}(x)$ points in a fixed direction in space and is static, but its magnitude varies in the $x$ direction, with a characteristic length scale $\lambda$ that is arbitrary. Similarly, $\vec{E}(t)$ points in a fixed direction in space and is spatially uniform, but its magnitude varies in time, with a characteristic time scale $\tau$ that is arbitrary. These electric and magnetic fields satisfy the homogeneous Maxwell equations, but not the inhomogeneous ones, so classically we should think of them as being supported by external currents. Within a quantum path integral they simply correspond to some particular vector potential $A_\mu$. Note that in the limits $\lambda\to\infty$ and $\tau\to\infty$ we regain the uniform field cases relevant for the Euler-Heisenberg-Schwinger effective action. We therefore expect that, for the inhomogeneous backgrounds (\ref{inhom}), the derivative expansion of the effective action should correspond to an expansion for large $\lambda$ and large $\tau$. We concentrate first on the magnetic field case, and then use a duality transformation $B\to iE$ to convert to the electric field case. 

Since each derivative of the magnetic field in (\ref{inhom}) produces a factor of $\frac{1}{\lambda}$, a natural dimensionless expansion parameter for the derivative expansion in the magnetic case is (restoring factors of $\hbar$ and $c$)
\begin{eqnarray}
{\hbar c\over eB\lambda^2} =\left({\sqrt{\hbar c/(eB)}\over \lambda}\right)^2
\label{maglength}
\end{eqnarray}
This dimensionless parameter is the square of the ratio of the magnetic length scale $\sqrt{\frac{\hbar c}{eB}}$ (which is set by the peak magnetic field magnitude $B\equiv|\vec{B}|$) to $\lambda$, the length scale of the spatial inhomogeneity of the magnetic field. Alternatively, we can combine this with the dimensionless perturbation parameter (\ref{realparam}) of the constant field case, to obtain another dimensionless expansion parameter
\begin{eqnarray}
\left({\hbar c\over eB\lambda^2}\right)\left( {eB\hbar\over m^2c^3}\right)=
{\hbar^2\over m^2 c^2 \lambda^2} = \left({\hbar/(mc)\over \lambda}\right)^2
\label{compton}
\end{eqnarray}
which is essentially (up to factors of $2\pi$) the square of the ratio of the Compton wavelength of the electron to the inhomogeneity scale $\lambda$. Thus, for a magenetic background with a macroscopic inhomogeneity scale $\lambda$, the ratio $\frac{1}{m^2\lambda^2}$ is extremely small. In this form, we clearly recognize the derivative expansion as an expansion in inverse powers of $m^2$, as in a general effective field theory expansion (\ref{eff}).

Indeed, these expectations are borne out by the exact renormalized effective action \cite{dh1}, which has the double perturbative expansion (we set $c$ and $\hbar$ to 1 again)
\begin{eqnarray}
S\sim -\frac{L^{2} \lambda T m^{4}}{8\pi ^{3/2}}
{\sum_{j=0}^{\infty}}\frac{1}{j!}
\left( \frac{1}{m^2\lambda^2}\right)^j\sum_{k=1}^{\infty }\frac{\Gamma
(2k+j)\Gamma (2k+j-2){\cal B}_{2k+2j}}{\Gamma (2k+1)\Gamma (2k+j+\frac{1}{2})}
\left( \frac{2eB}{m^{2}}\right)^{2k}
\label{exact}
\end{eqnarray}
In (\ref{exact}) it is understood that the double sum excludes the $(j=0, k=1)$ term, as this term contributes to the logarithmically divergent charge renormalization of the bare action \cite{schwinger,dh1}. 

We emphasize that the summation indexed by $j$ in (\ref{exact}) corresponds precisely to the orders of the derivative expansion. This has been verified \cite{dh1} by comparison with independent derivative expansion calculations of the leading and first-correction term, computations that were done using the proper-time method \cite{daniel,shovkovy}. For example, to make the comparison with the leading term of the derivative expansion we simply take the Euler-Heisenberg constant field answer (\ref{eh}), replace $B$ by $B(x)=B\, {\rm sech}^2(\frac{x}{\lambda})$ and do the $x$ integrals, using the fact that $\int_{-\infty}^\infty\, {\rm sech}^{4n}(x)dx =\sqrt{\pi}\Gamma(2n)/\Gamma(2n+1/2)$. This reproduces the $j=0$ term in (\ref{exact}). A similar argument \cite{dh1} holds for the first correction term in the derivative expansion, which reproduces the $j=1$ term in (\ref{exact}).

It is important to note that the perturbative expression (\ref{exact}) for the effective action is an explicit double sum (with no remaining integrals) in terms of two dimensionless parameters. One parameter $\frac{eB}{m^2}$ characterizes the perturbative expansion in powers of the coupling $e$, while the other parameter $\frac{1}{m^2\lambda^2}$ characterizes the derivative expansion. Moreover, all the expansion coefficients are known exactly. Thus, we can apply to this effective action the standard techniques for the analysis of divergent series (such as Borel summation). 

\subsection{Leading Order in Derivative Expansion}

Consider a fixed order $j$ of the derivative expansion. This still involves a perturbative expansion in powers of the coupling $e$. For $j=0$, from (\ref{exact}) we see that the perturbative expansion in the magnetic case is
\begin{eqnarray}
S^{(j=0)}\sim -\frac{L^{2}\lambda T m^{4}}{8\pi ^{3/2}}
\left( \frac{2eB}{m^{2}}\right)^4 \sum_{k=0}^{\infty} \frac{\Gamma(2k+4)\Gamma (2k+2){\cal B}_{2k+4}} {\Gamma (2k+5)\Gamma (2k+\frac{9}{2})}
\left(\frac{2eB}{m^{2}}\right)^{2k} \qquad\qquad (\frac{eB}{m^2}\to 0^+)
\label{bzero}
\end{eqnarray}
The expansion coefficients alternate in sign and grow factorially with $k$:
\begin{eqnarray}
\frac{\Gamma(2k+4)\Gamma (2k+2){\cal B}_{2k+4}}
{\Gamma (2k+5)\Gamma (2k+\frac{9}{2})}\sim {2(-1)^{k+1}\over (2\pi)^{2k+4}} \Gamma(2k+\frac{3}{2})\left[1+{\rm O}\left(\frac{1}{k}\right)\right] \qquad\qquad, k\to \infty
\label{bgrowth}
\end{eqnarray}
Thus, we are in the situation described by (\ref{general}). Therefore, this leading order ($j=0$) in the derivative expansion is a divergent but Borel summable series [this is hardly surprising, as we already showed in the previous section that the Euler-Heisenberg effective action is itself a divergent but Borel summable series]. Applying the Borel summation formula (\ref{genborel}), the leading Borel approximation for the series (\ref{bzero}) is
\begin{eqnarray}
S^{(j=0)}\sim \frac{L^{2}\lambda T m^{4}}{4\pi^{11/2}}
\left( \frac{eB}{m^{2}}\right)^{5/2} \int_0^\infty ds\,
{\sqrt{s} \over 1+s^2/\pi^2} \exp\left[-\frac{m^2}{eB} s\right] \qquad\qquad 
(\frac{eB}{m^2}\to 0^+)
\label{bzb}
\end{eqnarray}
It is instructive to compare this with the leading Borel approximation (\ref{firstborel}) to the effective action for a {\it uniform} magnetic background $B$. To make this comparison we replace the uniform background $B$ in (\ref{firstborel}) with the inhomogeneous background 
$B(x)=B\,{\rm sech}^2(x/\lambda)$, and then perform the $x$ integration. Thus
\begin{eqnarray}
S_{\rm leading}&\sim& {e^2 \over 4\pi^6} \int_0^\infty \, ds 
\left({s\over 1+s^2/\pi^2}\right)\,\int d^4x\, B(x)^2 \, e^{-m^2 s/(eB(x))} \nonumber\\
&=&\frac{L^{2}\lambda T e^2 B^2}{4\pi^{11/2}}
\int_0^\infty ds\, {s \over 1+s^2/\pi^2} \exp\left[-\frac{m^2}{eB} s\right]\, \Psi(\frac{1}{2},-1;\frac{m^2}{eB} s) 
\label{bzbcheck}
\end{eqnarray}
where $\Psi(a,b;z)$ is the confluent hypergeometric function, with integral representation
\begin{eqnarray}
\Psi(a,b;z)=\frac{1}{\Gamma(a)}\int_0^\infty  e^{-zt} t^{a-1} (1+t)^{b-a-1} dt
\label{chyper}
\end{eqnarray}
Noting that $\Psi(a,b;z)\sim z^{-a}$ for large $z$, we see that (\ref{bzbcheck}) is indeed in agreement with (\ref{bzb}) when the perturbative expansion parameter $\frac{eB}{m^2}$ is small.

Now consider the $j=0$ term of the derivative expansion for the inhomogeneous electric background $E(t)=E\,{\rm sech}^2(t/\tau)$. Perturbatively, we replace $B^2$ by $-E^2$ in the expansion (\ref{bzero}), so that the expansion coefficients are now non-alternating. Thus, in the electric case the series is divergent but {\it not} Borel summable. Nevertheless, we can use the Borel dispersion relations (\ref{genimag}) to compute the imaginary part of the effective action. A direct application of (\ref{genimag}) leads to 
\begin{eqnarray}
{\rm Im}S^{(j=0)}\sim {L^3 \tau m^4\over 8 \pi^3} \left(\frac{eE}{m^2}\right)^{5/2} \exp\left[-\frac{m^2\pi}{eE}\right]
\qquad\qquad (\frac{eE}{m^2}\to 0^+)
\label{ezb}
\end{eqnarray}
Now compare this with the leading order Borel result (\ref{leadingimag}) for a uniform electric background $E$. Replacing $E$ by the inhomogeneous field $E(t)=E\,{\rm sech}^2(t/\tau)$ and then performing the $t$ integration, we find
\begin{eqnarray}
{\rm Im}S_{\rm leading}\sim {L^3 \tau e^2 E^2\over 8 \pi^{5/2}}  \exp\left[-\frac{m^2\pi}{eE}\right]\, \Psi(\frac{1}{2},-1;\frac{m^2 \pi}{eE})
\qquad\qquad (\frac{eE}{m^2}\to 0^+)
\label{ezbcheck}
\end{eqnarray}
When the perturbative parameter $\frac{eE}{m^2}$ is small, this agrees precisely with the resummed answer in (\ref{ezb}). 

\subsection{First Correction in Derivative Expansion}

A similar analysis for the $j=1$ term (i.e. the first derivative expansion correction term) in (\ref{exact}) shows that in the magnetic case the perturbative expansion in powers of $(\frac{2eB}{m^2})^2$ is Borel summable and has a leading Borel approximation:
\begin{eqnarray}
S^{(j=1)}\sim \frac{L^{2}\lambda T m^{4}}{16\pi^{11/2}}
\left( \frac{eB}{m^{2}}\right)^{5/2} \left({m^4\over e^3 B^3 \lambda^2}\right) \int_0^\infty \frac{ds}{s}\,\left({s^{5/2} \over 1+s^2/\pi^2}\right)\, \exp\left[-\frac{m^2}{eB} s\right]
\label{boneb}
\end{eqnarray}
We can compare this with the first correction in the derivative expansion which has been computed independently using proper-time methods \cite{daniel,shovkovy}:
\begin{eqnarray}
S_{\rm first}\sim -{1\over 64\pi^2} \int_0^\infty \frac{ds}{s} 
\left[s({\rm coth}(s)-\frac{1}{s}-\frac{s}{3})\right]^{\prime\prime\prime}\, \int d^4x \, {(e B^\prime(x))^2\over eB(x)} \, 
\exp\left[ -\frac{m^2}{eB(x)}s\right]
\label{fo}
\end{eqnarray}
With the inhomogeneous background $B(x)=B{\rm sech}^2(\frac{x}{\lambda})$, the space-time integrals can be done to yield
\begin{eqnarray}
S_{\rm first}\sim -{L^2\lambda T\over 32 \pi^{3/2}}\left(\frac{eB}{\lambda^2}\right) \int_0^\infty \frac{ds}{s} 
\left[s({\rm coth}(s)-\frac{1}{s}-\frac{s}{3})\right]^{\prime\prime\prime} \, \exp\left[ -\frac{m^2}{eB}s\right]\, \Psi(\frac{3}{2},0;\frac{m^2}{eB}s)
\label{fd}
\end{eqnarray}
For small perturbative parameter $\frac{eB}{m^2}$ this reduces (after some integrations by parts in $s$) to the expression (\ref{boneb}) which was obtained by Borel summation of the $j=1$ term of the double series (\ref{exact}).

In the electric case, the $j=1$ term in (\ref{exact}) is a perturbative series expansion that is divergent but not Borel summable, as the expansion coefficients are non-alternating. We can compute the imaginary contribution to the effective action using the Borel dispersion relation result  (\ref{genimag})
\begin{eqnarray}
{\rm Im}S^{(j=1)} \sim {L^3\tau m^4 \over 32 \pi^2} \left(\frac{1}{m^2\tau^2}\right) \left(\frac{eE}{m^2}\right)^{-1/2}\, \exp\left[-\frac{m^2\pi}{eE}\right]  \qquad\qquad (\frac{eE}{m^2}\to 0^+)
\label{efe}
\end{eqnarray}
This should be compared to the first-order derivative expansion result from a field-theoretic calculation \cite{ted,shovkovy} 
\begin{eqnarray}
{\rm Im} S_{\rm first} \sim {m^6\over 64\pi}\int d^4x\, {(\partial_0 E(t))^2\over E(t)^4} \, \exp\left[-\frac{m^2\pi}{eE(t)}\right]
\label{efefirst}
\end{eqnarray}
With the inhomogeneous background $E(t)=E{\rm sech}^2(\frac{t}{\tau})$, the space-time integrals can be done to yield
\begin{eqnarray}
{\rm Im} S_{\rm first} &\sim& {L^3 m^6\over 8\pi \tau e^2 E^2} \int_1^\infty dz\sqrt{z^2-1}\, z^2 \, \exp\left[-\frac{m^2\pi}{eE}z^2\right]
\qquad\qquad (\frac{eE}{m^2}\to 0^+)\nonumber\\
&=&  {L^3 m^6\over 32\sqrt{\pi} \tau e^2 E^2} \exp\left[-\frac{m^2\pi}{eE}\right]\,\Psi(\frac{3}{2},3;\frac{m^2\pi}{eE}) 
\label{efecheck}
\end{eqnarray}
which agrees precisely with (\ref{efe}) when $\frac{eE}{m^2}$ is small.

\subsection{Resumming the Derivative Expansion}

Having verified explicitly that the Borel techniques work for the first two orders of the derivative expansion, for both the magnetic and electric background, we now turn to the higher orders $j\geq 2$ of the derivative expansion. These are very difficult to compute with field theory techniques \cite{schubert}. Nevertheless, for the particular inhomogeneous backgrounds in (\ref{inhom}), the exact result (\ref{exact}) contains {\it all} orders in the derivative expansion, and so it is a simple matter to study the divergence properties of each order $j$ of the derivative expansion.

From (\ref{exact}), the $j^{\rm th}$ order derivative expansion contribution to the effective action is ($j\geq 1$)
\begin{eqnarray}
S^{(j)}\sim -\frac{L^{2}\lambda T m^{4}}{8\pi ^{3/2}}
\frac{1}{j!}\left( \frac{1}{m^2\lambda ^{2}}\right)^j \left( \frac{2eB}{m^{2}}\right)^2\sum_{k=0}^{\infty} \frac{\Gamma(2k+j)\Gamma (2k+j+2){\cal B}_{2k+2j+2}}{\Gamma (2k+3)\Gamma (2k+j+\frac{5}{2})}
\left(\frac{2eB}{m^{2}}\right)^{2k}
\label{fixedj}
\end{eqnarray}
For fixed $j$ this contribution is itself a perturbative series expansion in terms of the dimensionless parameter $g=\frac{4 e^2B^2}{m^4}$, with expansion coefficients
\begin{eqnarray} 
a_k^{(j)}&=&\frac{\Gamma(2k+j) \Gamma(2k+j+2){\cal B}_{2k+2j+2}}
{\Gamma (2k+3)\Gamma (2k+j+\frac{5}{2})}\nonumber\\
&\sim & 2 (-1)^{j+k} {\Gamma(2k+3j-\frac{1}{2})\over (2\pi)^{2j+2k+2}} \qquad\qquad (k\to\infty;\,\,j\,\, {\rm fixed})
\label{dc}
\end{eqnarray}
These coefficients alternate in sign and grow factorially with $k$, for any fixed $j\geq 1$. Therefore, for the inhomogeneous magnetic background in (\ref{inhom}), each order of the derivative expansion is a divergent but Borel summable asymptotic series. Using the leading large $k$ behavior in (\ref{dc}) together with the Borel integral (\ref{borel}), the leading Borel approximation to the $j^{\rm th}$ order of the derivative expansion is (for $j\geq 1$)
\begin{eqnarray}
S^{(j)}\sim -
{L^2\lambda T m^4\over 4\pi^{7/2}}\left(\frac{eB}{m^2}\right)^{5/2} \, \int_0^\infty \frac{ds}{s^{3/2}} 
{\exp\left[-\frac{m^2}{eB}s\right]\over 1+s^2/\pi^2} \, \frac{1}{j!}\, \left({-m^4 s^3\over 4\pi^2 \lambda^2 e^3 B^3}\right)^j 
\qquad\qquad (\frac{eB}{m^2}\to 0^+)
\label{jb}
\end{eqnarray}
Note the remarkable fact that these leading Borel approximations, for each order $j$ of the derivative expansion, can be resummed into an exponential. The $j=0$ term must be treated separately [because the sum in (\ref{exact}) begins at $k=2$ when $j=0$, because of charge renormalization]. Combining the $j=0$ result (\ref{bzero}) with the $j\geq 1$ result (\ref{jb}) we find
\begin{eqnarray}
S\sim -{L^2 \lambda T m^4\over 2\pi^3} \left(\frac{eB}{m^2}\right)^2 -
{L^2\lambda T m^4\over 4 \pi^{7/2}} \left(\frac{eB}{m^2}\right)^{5/2} 
\int_0^\infty \frac{ds}{s^{3/2}} {1\over 1+s^2/\pi^2}\, 
\exp\left[ -\frac{m^2 s}{eB}\left\{1+ \left(\frac{m}{eB\lambda}\right)^2
\frac{s^2}{4\pi^2} \right\} \right]
\label{bresum}
\end{eqnarray}
The first term is a finite charge renormalization, on top of the usual infinite charge renormalization for the uniform field case \cite{schwinger}. In the second term, we see the interesting result that the leading Borel approximations to each order of the derivative expansion {\it exponentiate} when they are resummed. 

A similar phenomenon occurs with the electric background. For any $j\geq 1$, a straightforward application of the Borel dispersion result (\ref{genimag}), using the growth estimate of the coefficients in (\ref{dc}) gives
\begin{eqnarray}
{\rm Im} S^{(j)}\sim {L^3\tau m^4\over 8\pi^3} \left(\frac{eE}{m^2}\right)^{5/2}\, \exp\left[-\frac{m^2\pi}{eE}\right] \, \frac{1}{j!}\, \left({m^4 \pi\over 4 \tau^2 e^3 E^3}\right)^j
\qquad\qquad (\frac{eE}{m^2}\to 0^+)
\label{ej}
\end{eqnarray}
In fact, comparing with (\ref{ezb}) we see that this result for the imaginary part also holds for $j=0$. Thus, resumming the derivative expansion, the result immediately exponentiates \footnote{Of course, the finite renormalization found in the magnetic case (\ref{bresum}) does not affect the imaginary part of the effective action in the electric case.} to
\begin{eqnarray}
{\rm Im}S\sim {L^3\tau m^4\over 8\pi^3} \left(\frac{eE}{m^2}\right)^{5/2}\, \exp\left[-\frac{m^2\pi}{eE}
\left\{1-\frac{1}{4} \left(\frac{m}{eE\tau}\right)^2\right\}\right] 
\qquad\qquad (\frac{eE}{m^2}\to 0^+)
\label{eresum}
\end{eqnarray}
It is very interesting to see that the leading Borel approximations to each order of the derivative expansion can be resummed into an exponentiated form. 
Thus, resumming the leading Borel contributions amounts to a resummed perturbative modification of the non-perturbative exponent $\frac{\pi m^2}{eE}$ in the Schwinger uniform field result (\ref{leadingimag}). Clearly, the modification of the exponent derived in (\ref{eresum}) is much more significant than a modification of the prefactor, which is all that is obtained by looking at a single (low) order of the derivative expansion \cite{shovkovy,dh2}. We stress that we are able to exponentiate the corrections because in this case we know the large order perturbative behavior for {\it every} order of the derivative expansion. Finally, notice the appearance of the dimensionless parameter $\frac{m}{eE\tau}$ in the correction to the exponent in (\ref{eresum}). We will address the significance of this parameter below.

\subsection{Divergence of the Derivative Expansion}

In the previous section we took the leading Borel approximation to the perturbative expansion [the sum over $k$ in (\ref{exact})] and then performed the derivative expansion [the sum over $j$ in (\ref{exact})] exactly. Actually, it is possible to re-sum the perturbative $k$ expansion in (\ref{exact}), and express it as an integral. That is, we can write the effective action as a single (derivative expansion) series:
\begin{eqnarray}
S\sim -\frac{L^2\lambda T m^4}{8\pi^{3/2}} 
\sum_{j=0}^\infty a_j \left( \frac{1}{m^2\lambda^2}\right)^j  
\qquad\qquad (\frac{1}{m^2\lambda^2}\to 0^+)
\label{dd}
\end{eqnarray}
where the expansion coefficients $a_j$  are now {\it functions} of the parameter $\frac{eB}{m^2}$. To study the divergence properties of this derivative expansion we need to know the rate of growth, for large $j$,  of the expansion coefficients $a_j$ appearing in (\ref{dd}). For $j\geq 3$ there is a simple integral representation for these coefficients [this amounts to summing the perturbative expansion in the double series (\ref{exact}), thereby reducing (\ref{exact}) to the single series in (\ref{dd})]:
\begin{eqnarray}
a_j= (-1)^{j+1}{\pi \Gamma(j-2)\over 2 j\Gamma(j+\frac{1}{2})}
\int_0^\infty \hskip-5pt {\rm cosech}^2(\pi s) s^{2j} 
\left[{}_2 F_1 (j,j-2;j+\frac{1}{2};\frac{2ieBs}{m^2}) +
{}_2F_1 (j,j-2;j+\frac{1}{2};-\frac{2ieBs}{m^2}) -2\right] ds
\end{eqnarray}
Here, ${}_2F_1(a,b;c;z)$ is the standard hypergeometric function \cite{gradshteyn}: 
\begin{eqnarray}
{}_2F_1(a,b;c;z)= {\Gamma(c)\over \Gamma(a)\Gamma(b)} \sum_{k=0}^\infty 
{\Gamma(a+k)\Gamma(b+k)\over \Gamma(c+k)} \frac{z^k}{k!}
\label{hyper}
\end{eqnarray}
For any $j$ it is straightforward to evaluate these coefficients $a_j$ 
numerically for various values of the dimensionless parameter $\frac{eB}{m^2}$.
We find that the coefficients $a_j$ alternate in sign and grow in magnitude (for large $j$) like $\alpha^j \Gamma(2j+\gamma)$, with some real $\alpha$, $\gamma$. This is illustrated in Figure 2, where the ratio of successive magnitudes $|a_{j+1}|/|a_j|$ shows a clear quadratic growth, for various values of $\frac{eB}{m^2}$. This shows that the derivative expansion itself (\ref{dd}) is a divergent but Borel summable asymptotic series.

\begin{figure}[htb]
\centering{\epsfig{file=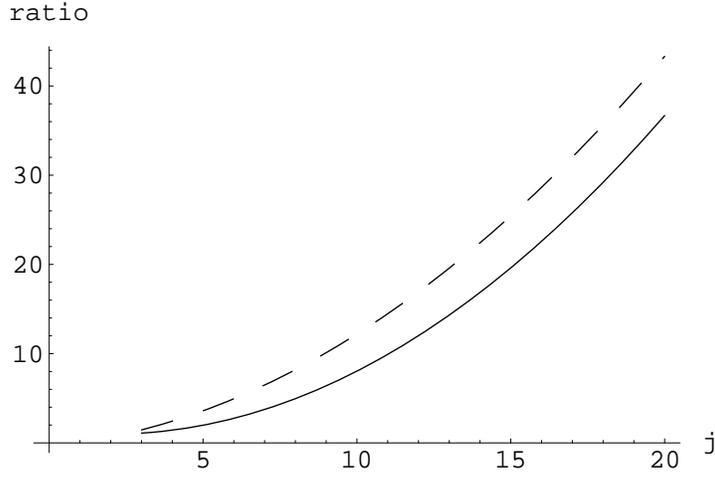, width = 4in, height=2.5in}}
\vskip .5cm
\caption{This figure plots the ratios $|a_{j+1}/a_j|$ of the successive expansion coefficients $a_j$ in the derivative expansion (\protect{\ref{dd}}), for two different values of the dimensionless expansion parameter $g$ defined in (\protect{\ref{param}}). The solid line refers to $g=0.1$ and the dashed line corresponds to $g=0.0001$. Note the quadratic growth of this ratio, indicating that $|a_j| \sim \alpha^j \Gamma(2j+\gamma)$ for large $j$.}
\end{figure}

\subsection{Resumming the Perturbative Expansion}

In this section we apply the leading Borel approximation to the derivative expansion, for a given order of the perturbative expansion, and then re-sum the perturbative expansion.

First, take a fixed order $k\geq 2$ of the perturbative expansion in (\ref{exact})
\begin{eqnarray}
S^{(k)}\sim -\frac{L^{2}\lambda T m^{4}}{8\pi ^{3/2}} 
\frac{1}{\Gamma(2k+1)}\left(\frac{2eB}{m^{2}}\right)^{2k} 
{\sum_{j=0}^{\infty}}
\frac{\Gamma(2k+j)\Gamma (2k+j-2){\cal B}_{2k+2j}}{\Gamma (j+1)\Gamma (2k+j+\frac{1}{2})} \left( \frac{1}{m^2\lambda^2}\right)^j 
\qquad\qquad (\frac{1}{m^2\lambda^2}\to 0^+)
\label{pert}
\end{eqnarray}
This is a derivative expansion in terms of the dimensionless parameter $\frac{1}{m^2\lambda^2}$, with expansion coefficients
\begin{eqnarray} 
a_j^{(k)}&=&\frac{\Gamma(j+2k) \Gamma(j+2k-2){\cal B}_{2j+2k}}
{\Gamma (j+1)\Gamma (j+2k+\frac{1}{2})} \nonumber\\
&\sim& 2 (-1)^{j+k+1} {2^{7/2-2k}\Gamma(2j+4k-\frac{5}{2})\over (2\pi)^{2j+2k}}
\qquad\qquad (j\to\infty; \,\, k\,\, {\rm fixed})
\label{pt}
\end{eqnarray}
Thus, the coefficients $a_j^{(k)}$ alternate in sign and grow factorially with $j$, for any fixed $k\geq 2$. Therefore, for any fixed order of perturbation theory in $\frac{eB}{m^2}$, the derivative expansion is a divergent but Borel summable asymptotic series. 

Now consider resumming the leading Borel summation of each order $k$ of the
perturbative expansion for the inhomogeneous electric background in
(\ref{inhom}). From (\ref{pt}) and the Borel dispersion relation (\ref{genimag})
the imaginary part of the effective action at order $k$ of the perturbative
expansion is
\begin{eqnarray}
{\rm Im}S^{(k)}\sim {L^3 m^{3/2}\over 4\pi^3 \tau^{3/2}}\, {(2\pi eE\tau^2)^{2k}
\over (2k)!} \, \exp^{-2\pi m\tau} \qquad\qquad (\frac{1}{m\tau}\to 0^+)
\label{kp}
\end{eqnarray}
This is in fact valid for all $k\geq 1$. Resumming these leading Borel
contributions gives the leading behavior, for large $eE\tau^2$ (as is appropriate for the derivative
expansion),
\begin{eqnarray}
{\rm Im}S\sim {L^3 m^{3/2}\over 8\pi^3 \tau^{3/2}}\, 
\exp\left[-2\pi m\tau\left(1-\frac{eE\tau}{m}\right)\right]
\qquad\qquad (\frac{1}{m\tau}\to 0^+)
\label{presum}
\end{eqnarray}
Once again, we see that the resummation of the leading Borel contributions exponentiates, producing an exponent that is modified from that found in the Schwinger uniform field result (\ref{leadingimag}).

However, the exponential behavior in (\ref{presum}) is very different from the exponential behavior in (\ref{eresum}), and indeed from the exponential behavior in the uniform case (\ref{leadingimag}). To understand this difference, we first recall that the exponential behavior in (\ref{eresum}) was obtained by resumming the leading Borel approximations to each order of the derivative expansion, while the exponential behavior in (\ref{presum}) was obtained by resumming the leading Borel approximations to each order of the perturbative expansion. That is, to obtain (\ref{eresum}) we take the leading Borel approximation for the $k$ summation in (\ref{exact}), for each fixed $j$, and then resum over $j$; while to obtain (\ref{presum}) we take the leading Borel approximation for the $j$ summation in (\ref{exact}), for each fixed $k$, and then resum over $k$.

The difference between these two approaches is
governed by the relative size of the two dimensionless expansion parameters. In the perturbative expansion we assume that $\frac{eE}{m^2}$ is small, and in the derivative expansion we assume that $\frac{1}{m\tau}$ is small. The distinction between the two answers (\ref{eresum}) and (\ref{presum}) depends on the dimensionless combination:
\begin{eqnarray}
\frac{eE\tau}{m}={(\frac{eE}{m^2})\over (\frac{1}{m\tau})}
\label{gov}
\end{eqnarray}
There are two natural regimes of interest:
\begin{eqnarray}
\matrix{{\rm non-perturbative~regime}~:\frac{eE\tau}{m}\gg 1 \cr\cr 
{\rm perturbative~regime}~: \frac{eE\tau}{m}\ll 1 }
\label{regimes}
\end{eqnarray}
In the non-perturbative regime, $\frac{eE\tau}{m}\gg 1$ implies that
\begin{eqnarray}
m\tau \gg \frac{m^2}{eE} \qquad\qquad \Rightarrow \qquad\qquad e^{-m^2\pi/(eE)} \gg e^{-2\pi m\tau}
\end{eqnarray}
Therefore, in this regime we expect the leading exponential contribution to ${\rm Im} S$ to be the Schwinger uniform field factor $\exp(-\frac{m^2\pi}{eE})$, as indeed is found in (\ref{eresum}). The resummation of leading Borel approximations derived in (\ref{eresum}) gives corrections to this leading exponent
\begin{equation}
\frac{m^2\pi}{eE}\quad\to\quad \frac{m^2\pi}{eE}\left[1-\frac{1}{4}\left(\frac{m}{eE\tau}\right)^2\right]
\label{bc1}
\end{equation}
which has the form of a small correction in terms of the parameter $\frac{m}{eE\tau}$, which is small in this non-perturbative regime.

On the other hand, in the perturbative regime, $\frac{eE\tau}{m}\ll 1$ implies that
\begin{eqnarray}
m\tau \ll \frac{m^2}{eE} \qquad\qquad \Rightarrow \qquad\qquad e^{-2\pi m\tau}\gg e^{-m^2\pi/(eE)}
\end{eqnarray}
In this regime, the exponential factor $e^{-2\pi m\tau}$ dominates the Schwinger factor $e^{-m^2\pi/(eE)}$ and gives a new leading contribution to ${\rm Im} S$. The resummation of leading Borel approximations in (\ref{presum}) gives corrections to this leading exponent
\begin{equation}
2\pi m\tau \quad\to\quad 2\pi m\tau\left[1-\frac{eE\tau}{m}\right]
\label{bc2}
\end{equation}
where in the perturbative regime the parameter $\frac{eE\tau}{m}$ is small.

Now ask the question: how does the time dependence of the inhomogeneous electric background in (\ref{inhom}) modify Schwinger's constant field result (\ref{leadingimag})? The answer depends critically on how the characteristic time scale $\tau$ of the inhomogeneity relates to the time scale $\frac{m}{eE}$ set by the peak electric field $E$. This then determines [see (\ref{gov})] the relative magintude of the two expansion parameters $\frac{1}{m\tau}$ (corresponding to the derivative expansion) and $\frac{eE}{m^2}$ (corresponding to the perturbative expansion).

In the non-perturbative regime, the smaller of the two parameters is the derivative expansion parameter: $\frac{1}{m\tau}\ll\frac{eE}{m^2}$. Thus, we apply the leading Borel approximation to the perturbative expansion [the sum in powers of $\frac{eE}{m^2}$], and then resum the derivative expansion [the sum in powers of $\frac{1}{m\tau}$] exactly. This is exactly what was done in deriving the result (\ref{eresum}).

In the perturbative regime, the smaller of the two parameters is the perturbative parameter: $\frac{eE}{m^2}\ll\frac{1}{m\tau}$. Thus, we apply the leading Borel approximation to the derivative expansion [the sum in powers of $\frac{1}{m\tau}$], and then resum the perturbative expansion [the sum in powers of $\frac{eE}{m^2}$] exactly. This is exactly what was done in deriving the result (\ref{presum}).

Note that, in each case, our ability to treat the remaining sum {\it exactly} relied on the fact that the leading Borel approximations came out in a form that could be exponentiated. It is not a priori obvious that this dramatic simplification had to occur. However, in the next section we will see that this exponentiation is very natural in terms of a WKB formulation.

\section{Relation to WKB Analysis}

In a general background, $F_{\mu\nu}=F_{\mu\nu}(\vec{x},t)$, the effective action is too complicated to permit such a detailed Borel analysis as has been done in the previous sections for the uniform background and for the special inhomogeneous backgrounds in (\ref{inhom}). Clearly, we do not know the spectrum of the Dirac operator for a general background, so some sort of approximate expansion method, such as the derivative expansion,  is required. But the formal derivative expansion (\ref{der}) is not a series expansion, because more and more (independent) tensor structures appear with each new order of the derivative expansion. This is because with more derivatives there are more indices to be contracted in various ways. Another way of saying this is that the inhomogeneity of a general background cannot be characterized by a single (or even a finite number of) scale parameter(s), such as the length scale $\lambda$ or the time scale $\tau$ in (\ref{inhom}). Another problem is that in general it is difficult to estimate the size of various terms in such a derivative expansion when the background field strength $F_{\mu\nu}$ is an arbitrary function of space-time. So, even if we could organize the derivative expansion into a sensible series, it would be difficult to estimate the magnitude of the coefficients at very high orders in the series, as is needed for a Borel analysis.

Nevertheless, it is still instructive to consider a further generalization of the particular inhomogeneous backrounds in (\ref{inhom}). In this section we relax the condition that we know the exact spectrum of the Dirac operator, but keep the restriction that the backgrounds only depend on one space-time coordinate. This has the effect of reducing the spectral problem to that of an ordinary differential operator. As is clear from (\ref{action}), the effective action is determined by the spectrum of the operator
\begin{equation}
m^2+D^2\hskip -11pt / \hskip 10pt =\left[m^2+D_\mu D^\mu\right] {\bf 1} +\frac{e}{2}F_{\mu\nu} \sigma^{\mu\nu}
\label{op}
\end{equation}
where $\sigma^{\mu\nu}\equiv \frac{i}{2}[\gamma^\mu,\gamma^\nu]$. If we restrict our attention to inhomogeneous backgrounds that point in a fixed direction in space (say, the $z$ direction) and depend on just one space-time coordinate, then the operator in (\ref{op}) can be diagonalized with a suitable gauge choice and a suitable Dirac basis. For example, the spatially inhomogeneous magnetic field
\begin{equation}
\vec{B}(x)=\hat{z}\, B\, f^\prime (\frac{x}{\lambda})
\label{genb}
\end{equation}
can be realized with the vector potential $\vec{A}=(0,B\lambda\, f(\frac{x}{\lambda}),0)$. Then, in the standard Dirac representation \cite{IZ} for the gamma matrices, the operator $m^2+D^2\hskip -11pt /\hskip 5pt$ is diagonal, with diagonal entries (appearing twice each on the diagonal):
\begin{equation}
m^2+D_\mu D^\mu\pm e B(x)=m^2-k_0^2+k_z^2-\partial_x^2+(k_y-eB\lambda f(\frac{x}{\lambda}))^2\pm eB f^\prime (\frac{x}{\lambda})
\label{bdiag}
\end{equation}

Similarly, the time-dependent electric field
\begin{equation}
\vec{E}(t)=\hat{z}\, E\, f^\prime (\frac{t}{\tau})
\label{gene}
\end{equation}
can be realized with the vector potential $\vec{A}=(0,0,E\tau\, f(\frac{t}{\tau}))$. Then, in the standard chiral representation \cite{IZ} for the gamma matrices, the operator $m^2+D^2\hskip -11pt /\hskip 5pt$ is diagonal, with diagonal entries (appearing twice each on the diagonal):
\begin{equation}
m^2+D_\mu D^\mu\pm i e E(t)=m^2+k_x^2+k_y^2+\partial_0^2+(k_z-eE\tau f(\frac{t}{\tau}))^2\pm i eE f^\prime (\frac{t}{\tau})
\label{ediag}
\end{equation}

Therefore, in each case (\ref{genb}) and (\ref{gene}), the spectrum is determined by a one-dimensional ordinary differential operator. However, note the appearance of the factors of $i$ in the electric case (\ref{ediag}). This shows immediately the fundamental difference between a magnetic background and an electric background. In the magnetic case, the eigenvalues of the associated ordinary differential operator are real, while for the electric case, the eigenvalues of the associated ordinary differential operator have an imaginary part. (Note that in the magnetic case the boundary condition for the ordinary differential operator is for solutions that decay at $x=\pm \infty$, while in the electric case we seek solutions going like $e^{\mp i\epsilon t}$ at $t=\pm \infty$, corresponding to the particle/antiparticle pair \cite{brezin,cornwall}. The reader is encouraged to check all this explicitly for the simple case of a uniform background.) 

Schwinger's uniform field case corresponds to choosing the function $f$ appearing in (\ref{genb}) and (\ref{gene}) to be $f(u)=u$, while the inhomogeneous backgrounds in (\ref{inhom}) correspond to choosing $f(u)=\tanh(u)$. It is well known that in each case the spectrum of the associated ordinary differential operator in (\ref{bdiag}) and (\ref{ediag}) is exactly solvable \cite{morse}. This explains why it is possible to compute the effective action exactly for these backgrounds. For the more general fields in (\ref{genb}) and (\ref{gene}) it is not possible to find the exact spectrum. However, we can still use a WKB approach to approximate the spectrum. For a time-dependent, but spatially uniform, electric background this leads to the following WKB expression for the imaginary part of the effective action \cite{baha,dh2,ted}
\begin{eqnarray}
{\rm Im}S=\left(\frac{L}{2\pi}\right)^3 \sum_{n=1}^\infty \frac{1}{n}\, \int d^3k\, e^{-n\pi\Omega}
\label{wkb}
\end{eqnarray}
where the WKB exponent is
\begin{eqnarray}
\Omega=\frac{2i}{\pi}\,\int_{TP}\, \sqrt{\mu^2+\phi^2(t)}\, dt
\label{omega}
\end{eqnarray}
Here $\mu^2=m^2+k_x^2+k_y^2$ and 
\begin{equation}
\phi(t)=k_z-eA_e(t)=k_z-eE\tau f(t/\tau)
\label{phi}
\end{equation}
for the electric field in (\ref{gene}). The integration in (\ref{omega}) is between the turning points of the integrand [this expression is somewhat symbolic - in practice, the evaluation of $\Omega$ requires careful phase choices, depending on the form of the function $f(u)$ \cite{brezin,baha,ted}]. For the constant field case, with $f(u)=u$, one finds 
\begin{equation}
\Omega=\frac{\mu^2}{eE}
\label{constomega}
\end{equation}
Then the momentum integrals in (\ref{wkb}) can be done [recall the density of states integral: $\int dk_z=E$] to yield the familiar Schwinger result (\ref{fullimag}).

In the inhomogeneous electric field $E(t)=E\,{\rm sech}^2(t/\tau)$, which corresponds to $f(u)=\tanh(u)$, we can also compute $\Omega$ exactly \cite{baha,dh2}:
\begin{equation}
\Omega=\tau\left(\sqrt{\mu^2+(eE\tau +k_z)^2}+ \sqrt{\mu^2+(eE\tau -k_z)^2} 
-2 eE\tau\right)
\label{exactomega}
\end{equation}
It is a straightforward, but somewhat messy, computation to check that by doing the momentum integrals in (\ref{wkb}) with this expression for $\Omega$, one arrives at the exact integral representation, derived in \cite{dh2}, for the effective action in this inhomogeneous background. The WKB expression gives the {\it exact} result in this case because the uniform WKB approximation gives the exact spectrum of the differential operators (\ref{bdiag}) and (\ref{ediag}) when $f(u)=\tanh(u)$ \cite{comtet}.

Now consider this WKB exponent $\Omega$ in (\ref{exactomega}) in the non-perturbative and perturbative limits (\ref{regimes}). In the non-perturbative limit we can expand $\Omega$ in inverse powers of $e$ as
\begin{equation}
\Omega_{\rm non-pert}=\frac{\mu^2}{eE}\left[1+\frac{k_z^2}{(eE\tau)^2} -\frac{1}{4}\left(\frac{\mu}{eE\tau}\right)^2+\dots \right]
\label{long}
\end{equation}
In the perturbative limit, we obtain instead
\begin{equation}
\Omega_{\rm pert}=2\mu\tau \left[1+\frac{1}{2}\frac{k_z^2}{\mu^2} -\frac{eE\tau}{\mu}+\dots \right]
\label{short}
\end{equation}
Doing the momentum trace over $k_z$, the momentum in the direction of the field, in the WKB expression (\ref{wkb}) modifies the prefactor, but not the exponent. The traces over the transverse momenta can be done by approximating $\mu=\sqrt{m^2+k_\perp^2}\approx m+\frac{k_\perp^2}{2m}+\dots$. This effectively replaces $\mu$ with $m$ in (\ref{long}) and (\ref{short}), and the $k_\perp$ integrals in (\ref{wkb}) contribute to the prefactor. Then comparing (\ref{long}) and (\ref{short}) with (\ref{bc1}) and (\ref{bc2}), we see that we regain exactly the Borel resummed results (\ref{eresum}) and (\ref{presum}) obtained for the two extreme limits (\ref{regimes}). Thus, this WKB analysis explains why we found two different expressions, (\ref{eresum}) and (\ref{presum}), by Borel resummation of the double series (\ref{exact}), depending on the relative size of the two dimensionless expansion parameters. 

To conclude this discussion of the WKB approach, it is instructive to compare the $E(t)=E\, {\rm sech}^2(t/\tau)$ case with the case of an oscillating electric field $E(t)=E\, \sin(\omega t)$ which was studied in detail using WKB methods by Br\'ezin and Itzykson \cite{brezin}. This is not an exactly solvable case, but WKB provides a semiclassical result. Here $f(u)=-\cos(u)$ and so the WKB exponent is
\begin{equation}
\Omega=\frac{2i}{\pi}\int_{\rm TP}\, 
\sqrt{ \mu^2+[k_z+\frac{eE}{\omega}\cos(\omega t)]^2} \, dt
\label{sinwkb}
\end{equation}
While this integral cannot be done in closed form, one can consider the non-perturbative and perturbative limits \cite{brezin}. In the non-perturbative regime, where$\frac{m\omega}{eE}\ll 1$,
\begin{equation}
\Omega_{\rm non-pert} \approx \frac{\mu^2}{eE}\left[1+\frac{1}{2}\left(\frac{\omega}{eE}\right)^2k_z^2-
\frac{1}{8}\left(\frac{\mu\omega}{eE}\right)^2+\dots\right]
\label{wkblow}
\end{equation}
which is clearly analogous to the non-perturbative limit (\ref{long}) of the $E(t)=E\, {\rm sech}^2(t/\tau)$ case. Thus, in this regime, the Schwinger pair production rate has a modified exponent
\begin{equation}
\frac{m^2\pi}{eE}\quad\to\quad \frac{m^2\pi}{eE}\left[1-\frac{1}{8}\left(\frac{m\omega}{eE}\right)^2 \right]
\label{lowmod}
\end{equation}
On the other hand, in the perturbative regime, where $\frac{m\omega}{eE}\gg 1$,
\begin{equation}
\Omega_{\rm pert} \approx \frac{4\mu}{\pi \omega}\, \log\left(\frac{2\mu\omega}{eE}\right)
\label{wkbhigh}
\end{equation}
which is very different from the perturbative limit (\ref{short}) of the $E(t)=E\, {\rm sech}^2(t/\tau)$ case. In this regime, the WKB pair production rate for an oscillating time-dependent electric background becomes \cite{brezin}
\begin{equation}
{\rm Im} S\sim \frac{e^2E^2}{32\pi}  \left(\frac{e^2E^2}{4m^2\omega^2}\right)^{2m/\omega}
\label{highmod}
\end{equation}
This is a perturbative expression, with the perturbative parameter $(eE/(m\omega))^2$ raised to a power $2m/\omega$ which is the number of photons of frequency $\omega$ required to match the pair creation energy $2m$. 

So, while the non-perturbative results (\ref{long}) and (\ref{wkblow}) are very similar for the cases $E(t)=E {\rm sech}^2(t/\tau)$ and $E(t)=E \sin(\omega t)$ respectively, the corresponding perturbative results are very different from one another for these two time-dependent electric backgrounds. This difference can be traced to the different perturbative limits (\ref{short}) and (\ref{wkbhigh}) of the WKB exponent $\Omega$. Physically, this is not so surprising if we recall that the perturbative limits, $\frac{eE\tau}{m}\ll 1$ and $\frac{eE}{m\omega}\ll 1$, can also be thought of as short-pulse and high-frequency limits (respectively), in which case the ${\rm sech}^2(t/\tau)$ and $\sin(\omega t)$ profiles of the electric background are significantly different.

\section{Conclusion}

In conclusion, we have shown that the QED effective action is a divergent double series for the inhomogeneous magnetic background 
$B(x)=B {\rm sech}^2(x/\lambda)$ and for the inhomogeneous electric background $E(t)=E {\rm sech}^2(t/\tau)$. In particular, we have also demonstrated that for these inhomogeneous background fields the derivative expansion is itself divergent. Borel summation techniques have been used to relate the rate of divergence to the non-perturbative imaginary part of the effective action, which determines the pair-production rate in the time-dependent electric background. Remarkably, the leading Borel approximations exponentiate to yield corrections to the familiar exponent appearing in the constant field case. These resummations can also be explained using a WKB analysis of the imaginary part of the effective action.

Finally, we conclude by considering the question of Borel summability of the perturbative effective action when the inhomogeneous background has the more general form in (\ref{genb}) or (\ref{gene}). The exponential form of the WKB expression (\ref{wkb}) for the imaginary part of the effective action is very suggestive of the exponential imaginary parts found from the Borel dispersion relation (\ref{genimag}). Indeed, writing
\begin{equation}
\sum_{n=1}\frac{1}{n} e^{-n\pi \Omega}
=\frac{1}{\pi\Omega}\sum_{n=1}\frac{1}{n^2} (n\pi\Omega) e^{-n\pi\Omega}
\label{arr}
\end{equation}
we can read the Borel dispersion relation backwards [with $\beta=2$, $\gamma=1$, and $\sqrt{\frac{1}{\alpha g}}=n\pi\Omega$] to obtain the corresponding asymptotic expansion of the real part of the effective action
\begin{equation}
S\sim -2 T\left(\frac{L}{2\pi}\right)^3 \sum_l {{\cal B}_{2l+2}\over (2l+2)(2l+1)} \int d^3k\, \left(\frac{2}{\Omega}\right)^{2l+1}
\label{back}
\end{equation}
where in the magnetic case $\Omega\to -i \Omega$. This expression is consistent with the result from the resolvent method (although to be strictly correct we need to specify carefully phase conventions for $\Omega$ \cite{dh2,ted}). For example, in the constant magnetic case $\Omega=\mu^2/(eB)$, and it is easy to verify [recalling that $\int dk_z=B$] that the expansion (\ref{back}) reproduces the Euler-Heisenberg expansion (\ref{eh}). 

Interestingly, the expansion (\ref{back}) is a reorganization of the usual perturbative expansion of the effective action, and this reorganized form already makes manifest the generic divergence properties of the effective action, since the Bernoulli numbers have leading behavior ${\cal B}_{2n}\sim (-1)^{n+1} 2 (2n)!/(2\pi)^n$. Thus, we see that in the magnetic case the expansion (\ref{back}) has coefficients that alternate in sign and grow factorially in magnitude. In this formal sense, the divergence properties of the effective action discussed in this paper, for the special cases with $B={\rm constant}$ and $B(x)=B\, {\rm sech}^2(x/\lambda)$, appear to extend to more general backgrounds. However, we caution that this argument is formal, as it neglects the possible appearance of other poles and/or cuts in the Borel plane which might invalidate the naive use of the Borel dispersion relation (\ref{genimag}). Nevertheless, it is suggestive to associate the non-perturbative WKB expression (\ref{wkb}) for the imaginary part of the effective action with the divergent expansion (\ref{back}) for the real part.

\vskip 1cm

\section{Acknowledgements:}
This work has been supported by the U.S. Department of Energy grant DE-FG02-92ER40716.00, and by the University of Connecticut Research Foundation. We thank Carl Bender for helpful discussions.


\begin{references}
\bibitem{bern} The Bernoulli numbers ${\cal B}_{2n}$ are defined  by
$x/(e^x-1)=\sum_{n=0}^\infty\, {\cal B}_n x^n/n!$. Thus, the Bernoulli numbers appear in the Taylor series expansion of the coth function:
$\coth(x)=\sum_{n=0}^\infty 2^{2n} {\cal B}_{2n} x^{2n-1}/(2n)!$, $(|x|<\pi)$.
\end{references}
\end{document}